\documentclass[11pt]{article}
\usepackage[margin=1in]{geometry}
\usepackage[style=alphabetic,backend=biber,doi=true,maxnames=99]{biblatex}
\bibliography{ref_ipco.bib}

\input{arxiv_headers}
\newenvironment{proofof}[1]{\begin{proof}[{Proof of #1}]}{\end{proof}}

\DeclareMathOperator{\poly}{poly}

\newcommand{\Rp}{\ensuremath{\mathbb R}_{\geq 0}}
\newcommand{\R}{\ensuremath{\mathbb R}}
\newcommand{\Z}{\ensuremath{\mathbb Z}}
\newcommand{\Zp}{\ensuremath{\Z_{\geq 0}}}

\newcommand{\Q}{\ensuremath{\mathbb Q}}
\renewcommand{\Pr}{\operatorname{\mathbb{P}}}
\newcommand{\E}{\operatorname{\mathbb{E}}}

\newcommand{\laci}[1]{\begingroup \em\textcolor{ForestGreen}{Laci: \endgroup}}

\newcommand{\opt}{\ensuremath{\mathrm{OPT}\xspace}}
\newcommand{\lpopt}{\ensuremath{\mathrm{LP}\text{-}\opt}\xspace}
\newcommand{\safe}{\mathscr{S}}
\newcommand{\unsafe}{\mathscr{U}}

\newcommand{\fgc}{\mathrm{FGC}}
\newcommand{\oneonefgc}{(1,1)\text{-}\fgc}
\newcommand{\pqfgc}{(p,q)\text{-}\fgc}
\newcommand{\ptwofgc}{(p,2)\text{-}\fgc}
\newcommand{\oneqfgc}{(1,q)\text{-}\fgc}
\newcommand{\twoqfgc}{(2,q)\text{-}\fgc}

\newcommand{\ecss}{\mathrm{ECSS}}
\newcommand{\twoecss}{2\text{-}\ecss}
\newcommand{\kecss}{k\text{-}\ecss}
\newcommand{\capkecss}{\mathrm{Cap}\text{-}k\text{-}\ecss}

\newcommand{\As}{\mathscr{A}} 
\newcommand{\C}{\mathscr{C}} 
\newcommand{\I}{\mathcal{I}} 
\newcommand{\Rs}{\mathscr{R}} 

\newcommand{\sm}{\setminus}

\newcommand{\IGNORE}[1]{}

\newcommand{\vq}{{\boldsymbol{q}}}

\newcommand{\vfgc}[1][\ensuremath{\vq}]{\ensuremath{#1}\text{-}\fgc}

\newcommand{\va}{\boldsymbol{a}}

\title{An O(log n)-Approximation Algorithm for \\  
(p,q)-Flexible Graph Connectivity \\ via Independent Rounding{\thanks{A preliminary version of this paper appeared in the proceedings of the 26th International Conference on Integer Programming and Combinatorial Optimization, 2025~\cite{IV25}.}}}

\author{
    Sharat Ibrahimpur\thanks{{\tt sharat.ibrahimpur@uwaterloo.ca}.
    Work done when the author was a postdoctoral researcher at the Research Institute for Discrete Mathematics, University of Bonn, Bonn, Germany. The author is also grateful for the support received from the Deutsche Forschungsgemeinschaft (DFG, German Research Foundation) – 537750605.}
\and  
    László A. Végh \thanks{{\tt lvegh@uni-bonn.de}. Hertz Chair in the Transdisciplinary Research Area Modelling, University of Bonn, Bonn, Germany.}
}

\date{}

\begin{document}

\maketitle

\begin{abstract}

In the Flexible Graph Connectivity (FGC) problem, we are given an undirected multigraph on $n$ vertices with nonnegative edge costs, where each edge is classified as either \emph{safe} or \emph{unsafe}. Given integer parameters $p$ and $q$, the goal in $(p,q)$-FGC is to purchase a minimum-cost set of edges such that the resulting spanning subgraph remains $p$-edge-connected after the removal of any set of up to $q$ unsafe edges.

Our main contribution is an $O(\log n)$-approximation algorithm based on independent rounding, improving the previous best approximation ratio of $O(q \log n)$. Central to our approach is a new linear programming formulation of feasible solutions that encodes knapsack cover inequalities as cut-capacity constraints. 
Unlike prior work, the capacity of an edge in a cut may depend on the partially purchased solution for this cut. We show that the resulting linear program admits a polynomial-time separation oracle. Scaling the fractional solution by $\Theta(\log n)$ and applying independent rounding yields a feasible integral solution with constant probability; here, we leverage the knapsack cover inequalities to obtain strong concentration bounds for the rounded solution relative to any given partial solution. 
A key ingredient in both separation and rounding is the use of Karger's bound on the number of near-minimum cuts.

We also extend the $(p,q)$-FGC problem to model more than two safety tiers and show that our results and techniques extend naturally to this setting, albeit with increased approximation ratios and running times that scale with the number of tiers. 
\end{abstract}


\section{Introduction} \label{intro}

We consider the \emph{Flexible Graph Connectivity} ($\fgc$) model of Adjiashvili, Hommelsheim, and M\"uhlenthaler \cite{AHM22}, introduced as a means to capture nonuniformity of edges in survivable network design problems. 
In this model, we are given an undirected graph $G = (V,E)$ on $n$ vertices, a partition of $E = \safe \cup \unsafe$ into \emph{safe} and \emph{unsafe} edges, respectively, and nonnegative edge costs $\{c_e\}_{e \in E}$. 
Safe edges are assumed to be reliable, whereas an unsafe edge may fail. 
As typical in survivable network design problems, we want to buy a cheap set of edges $F \subseteq E$ such that the subgraph $(V,F)$ has some specified edge-connectivity properties, and these properties are robust against certain failures among unsafe edges. 

Formally, in an instance of the $\pqfgc$ problem, we are given an edge-connectivity parameter $p \in \Z_{\geq 1}$ and a robustness parameter $q \in \Z_{\geq 0}$. 
A subset $F \subseteq E$ is feasible if for any set $F'$ consisting of at most $q$ unsafe edges, the subgraph ${(V,F \sm F')}$ is $p$-edge-connected, i.e., for any pair of distinct vertices $u, v \in V$, there are $p$ edge-disjoint $(u,v)$-paths in $(V,F \sm F')$.  
The algorithmic goal in $\pqfgc$ is to find a feasible solution $F$ that minimizes $c(F) := \sum_{e \in F} c_e$. 
We assume that $E$ is feasible, which can be checked efficiently (see \Cref{feascheck}).  
Throughout the paper, let $\opt := c(F^*)$ denote the cost of an optimal solution for this instance. 
Note that this problem is already APX-hard \cite{GabowGTW09} as the classical problem of finding the smallest $2$-edge-connected spanning subgraph (abbreviated as unit-cost/unweighted $\twoecss$) of an undirected graph $G$ arises as a special case of $\oneonefgc$ where every edge is unsafe and has unit cost. 
Thus, it is natural to seek approximation algorithms for this problem that can produce a solution $F$ whose cost is at most $\alpha \cdot \opt$ for some small $\alpha > 1$, or establish complementary hardness-of-approximation results.

In the work \cite{AHM22} that pioneered this model, Adjiashvili et al. gave a $2.527$-approximation algorithm for $\oneonefgc$. 
Observe that $\oneonefgc$ already captures prominent survivable network design problems such as the $\twoecss$ problem, the tree augmentation problem (TAP), and the more general forest augmentation problem (FAP); note that $(1,0)$-$\fgc$ is the minimum cost spanning tree problem, which can be solved in polynomial time. 
In some sense, $\pqfgc$ can be viewed as an interpolation between the $p$-$\ecss$ problem (when all edges are safe) and the $(p+q)$-$\ecss$ problem (when all edges are unsafe). 

The FGC model attracted considerable interest in the network design community because of its elegance in capturing the nonuniformity of edges that appear in network design applications. 
Boyd, Cheriyan, Haddadan, and Ibrahimpur \cite{BCHI24} gave several approximation algorithms for different choices of $p$ and $q$. 
Their approximation ratios were as follows: $4$ when $q=1$, $q+1$ when $p=1$, and $O(q \log n)$ for general $p$ and $q$.  
Subsequently, two independent groups obtained improved approximation algorithms when $p$ or $q$ is small. 
Bansal, Cheriyan, Grout, and Ibrahimpur \cite{BCGI24} gave an $O(1)$-approximation when $q=2$. 
Concurrently, Chekuri and Jain \cite{CJ25} gave $O(p)$-approximations for the parameter settings $(p,2)$, $(p,3)$, and $(2p,4)$, and an $O(q)$-approximation when $p=2$. 
Bansal \cite{Bansal25} removed the dependence on $p$ in the approximation ratio for the $q=3$ case, thus obtaining a constant approximation ratio. 
Nutov \cite{Nutov25} gave improved approximation algorithms and/or analyses for $(p,1)$-/$(p,2)$-/$(2p,3)$-FGC. 
Recently, Bansal, Cheriyan, Khanna, and Simmons \cite{BCKS25} gave an $8$-approximation algorithm when $p=1$, thus removing the dependence on $q$ for the simplest ($1$-)connectivity case in the FGC model. 
The table below lists the current best approximation ratios for different regimes of $\pqfgc$ in the weighted setting.  

\begin{table}[ht]
\centering
\small
\caption{Best known approximation ratios for weighted versions of $\pqfgc$.
The values of cells containing $\rightarrow$, $\downarrow$, or both can be inferred by following one of the indicated directions until a compatible cell is found. For instance, the inferred values in the cells corresponding to $(1,3)$-, $(2,2)$-, $(3,3)$-, and $(3,4)$-FGC are $4, 6, 11+\varepsilon,$ and $O(\log n)$, respectively.}
\label{tab:fgc-2d}

\setlength{\tabcolsep}{4pt}
\renewcommand{\arraystretch}{1.15}
\begin{tabular}{|c|c|c|c|c|c|c|}
\hline
$p \backslash q$ & $0$ & $1$ & $2$ & $3$ & $4$ & general $q$ \\
\hline
$1$ & $1$ [Bor] & $\rightarrow$ & $\rightarrow$  & $\rightarrow$ & $\rightarrow$ & $\min(q+1,8)$ [BCHI,BCKS] \\ \hline
$2$ & $\downarrow$ & $\downarrow$ &  $\downarrow\,/\,\rightarrow$ & $\rightarrow$ & $\rightarrow$ & $2q+2$ [CJ] \\ \hline
$3$ & $\downarrow$ & $\downarrow$ &  $\downarrow$ & $\downarrow$ & $\downarrow\,/\,\rightarrow$ & $\downarrow$ \\ \hline
even $p$ & \multirow{2}{*}{$2$ [KV]}
& $4$ [BCHI] & $6$ [BCGI] & $12$ [Ban] & $6p+4$ [CJ]
& \multirow{2}{*}{$O(\log n)$ [IV]} \\ \cline{1-1} \cline{3-6} 
odd $p$ &  & $3.5+\varepsilon$ [Nut] & $7+\varepsilon$ [Nut] & $11+\varepsilon$ [Ban] & $\rightarrow$ &  \\ \hline
\end{tabular}

\vspace{0.5ex}
\footnotesize
\begin{tabular}{ll@{\qquad}ll@{\qquad}ll}
Bor & \cite{Boruvka} &
KV & \cite{KV94} &
BCHI & \cite{BCHI24} \\ 
BCGI & \cite{BCGI24} &  
CJ & \cite{CJ25} &
Nut & \cite{Nutov25} \\ 
Ban & \cite{Bansal25} &
BCKS & \cite{BCKS25} & 
IV & This work
\end{tabular}
\end{table}
\subsection{Our Contributions}

The focus of this work is $\pqfgc$ when $p$ and $q$ are arbitrary integer parameters. 
As mentioned above, the current best approximation algorithm for the general case has an approximation ratio of $O(q \log n)$ and is due to Boyd et al. \cite{BCHI24}. 
Our headline result is an improved $O(\log n)$-approximation algorithm. 
We note that the $O(q \log n)$-approximation algorithm is deterministic, whereas our algorithm is randomized.

\begin{theorem} \label{logapx}
Let $\I = (G=(V,E=\safe \cup \unsafe),c,p,q)$ be an instance of the $\pqfgc$ problem. 
There is a randomized algorithm that outputs, with constant probability, a feasible solution whose cost is $O(\log n)$ times the optimum value.  
The running time of the algorithm is polynomial in the input encoding size.
\end{theorem}

Our algorithm in \Cref{logapx} is based on a new linear integer programming (IP) formulation for $\pqfgc$. 
We present this formulation in \Cref{formulation} and also give an efficient separation oracle for the corresponding linear relaxation. 
The algorithm is then a standard independent randomized rounding. 
We solve the linear relaxation to obtain an optimal fractional solution $x$ and then independently include each edge $e \in E$ in the final solution with probability $\min(1,(100 \log n) \cdot x_e)$. 

We make two remarks here. 
First, Chekuri and Jain \cite{CJ25} also gave a valid IP formulation for $\pqfgc$ along with an efficient separation oracle for the associated linear program (LP). 
However, their approximation algorithms for $\ptwofgc$ and $\twoqfgc$ are not based on this formulation. 
Second, our LP is based on writing knapsack cover inequalities for some careful choice of edge capacities that may depend on the cut. 
We will elaborate on this shortly.  
Although our LP is a bit more involved, it was inspired by the work of Bansal, Cheriyan, Khanna, and Simmons \cite{BCKS25} in which they give an $O(1)$-approximation algorithm for $\oneqfgc$ for a general $q$.  

In \Cref{extension}, we give an extension of the $\pqfgc$ model to $\ell$ safety tiers for some integer $\ell \geq 2$. 
Formally, we are given an undirected multigraph $G = (V,E)$ with nonnegative edge costs $\{c_e\}_{e \in E}$, a partition $E = \cup_{i=1}^{\ell} E_i$ of the edge set into $\ell$ tiers of decreasing safety over $i \in [\ell]$, and an $\ell$-tuple $\vq = (q_1,\ldots,q_\ell) \in \Zp^{\ell}$ that encodes robustness parameters. 
The algorithmic goal is to find a minimum cost $F \subseteq E$ so that the spanning subgraph $(V,F \sm F')$ is connected for every choice of $F' \subseteq E$ satisfying $|F' \cap (E_1 \cup \ldots \cup E_i)| \leq q_1 + \ldots + q_i$ for all $i \in [\ell]$. 
We call this problem $\vfgc$, where the length of the robustness tuple $\vq$ implicitly encodes $\ell$, the number of safety tiers. 
Observe that $\pqfgc$ corresponds to taking $\ell = 2$, $q_1 = p-1$, $q_2 = q$, $E_1 = \safe$, and $E_2 = \unsafe$. 

Our results and techniques on $\pqfgc$ extend naturally to $\vfgc$, albeit with an additional $O(\ell^2)$ factor in the approximation ratio and an exponential dependence on $\ell$ in the running time. 

\begin{theorem} \label{logapx-ext}
Let $\I = (G=(V,E_1 \cup \ldots \cup E_\ell),c,\vq)$ be an instance of the $\vfgc$ problem with $\ell \geq 2$ safety tiers.  
There is a randomized algorithm that outputs, with constant probability, a feasible solution whose cost is $O(\ell^2 \log n)$ times the optimum value.  
The running time of the algorithm is polynomial in $n^\ell$ and the input encoding size. 
\end{theorem}

\subsection{Technical Overview and Comparison with Prior Work}

We now give an overview of our results on $\pqfgc$ with a brief discussion of relevant prior work. 
Our work builds on the techniques and ideas developed in a sequence of prior works \cite{BCKS25,BCHI24,CCKK15,CJ25}. 

The following notation and terminology will be used throughout the paper. 
The graphs we consider are connected and undirected, without self-loops. 
We allow the input graph $G$ to have parallel edges; however, any solution $F$ can contain at most one copy of any edge $e$ that appears in $E$. 
Any vertex set $R$ induces a \emph{cut} ${\delta(R) := \{ e \in E : |R \cap e| = 1 \}}$ consisting of all edges with exactly one endpoint in $R$. 
We call a vertex set $R$ \emph{nontrivial} if $R$ is neither empty nor all of $V$.  
Note that $\delta(\cdot)$ is symmetric, that is, $\delta(R) = \delta(V \sm R)$, and satisfies $\delta(R) = \emptyset$ if and only if $R$ is trivial, since we assume that $G$ is connected. 
For an edge set $F$, we define $\delta_F(R) := F \cap \delta(R)$ as the set of $F$-edges in the cut $\delta(R)$. 
For any real number $z \in \R$, we use $z^+$ to denote $\max(0, z)$. 
For any real-valued function $w : X \to \R$ on an arbitrary ground set $X$ and $Y \subset X$, we use $w(Y)$ to denote $\sum_{x \in Y} w(x)$. 
We use the same $w(Y)$ notation when $\{w_x\}_{x \in X}$ are some real numbers associated with elements of $X$. 

We say that an algorithm is \emph{efficient} if its running time is polynomial in the input encoding size. 
In the context of $\pqfgc$, the runtime of an efficient algorithm has a polynomial dependence on the encoding size of the edge costs. 
Note that in any nontrivial instance, $n$, $p$, and $q$ are bounded by $|E|$. 

\medskip

We start with the integer programming formulation of $\pqfgc$ that admits an efficient separation oracle for the associated linear relaxation.  
The following characterization of feasible solutions follows immediately from the problem statement.  

\begin{proposition} \label{cutchar-prop}
An edge set $F \subseteq E$ is feasible for $\pqfgc$ if and only if, for every $\emptyset \neq R \subsetneq V$, the edge set $\delta_F(R)$ contains $p$ safe edges or $p+q$ (safe or unsafe) edges in total. 
\end{proposition}

In \cite{BCHI24}, Boyd et al. observed a useful connection between $\pqfgc$ and the capacitated version of the $\kecss$ problem, denoted $\capkecss$. 
In an instance of the $\capkecss$ problem, we have a graph $G = (V,E)$ with nonnegative edge costs $\{c_e\}_{e \in E}$, integer edge capacities $\{u_e\}_{e \in E}$, and a global edge-connectivity parameter $k \in \Z_{\geq 1}$. 
A feasible solution here is an edge set $F \subseteq E$ such that for every $\emptyset \neq R \subsetneq V$ we have $u(\delta_F(R)) \geq k$, i.e., the global minimum cut in the capacitated graph $(V,F,\{u_e\}_{e \in F})$ is at least $k$. 
The algorithmic goal is then to find a feasible solution with the minimum cost. 
For the given instance of $\pqfgc$, one can construct an instance of the $\capkecss$ problem such that feasible solutions for the former problem remain feasible for the latter problem: we take $k = p(p+q)$, $u_e = p+q$ for safe edges and $u_e = p$ for unsafe edges. 
Although the two problems are not equivalent in general, the covering constraints in the $\capkecss$ problem are valid for $\pqfgc$ and therefore the optimum value for this instance of $\capkecss$ is at most $\opt$. 
Furthermore, Boyd et al. observed that a feasible solution for $\capkecss$ is not ``far'' from feasible for $\pqfgc$. 
In fact, they show that applying at most $q$ rounds of augmenting steps is sufficient to achieve $\pqfgc$ feasibility, where in each step a suitable instance of the hitting-set problem is solved to fix cuts that are still violated. 
Overall, this leads to an $O(q \log n)$-approximation guarantee.

By the above discussion, a natural set of valid inequalities to include in an IP formulation for $\pqfgc$ is the following: 
\begin{flalign} \label{eq:basiccapineq}
  \quad (p+q) \cdot x(\delta(R) \cap \safe) + p \cdot x(\delta(R) \cap \unsafe) \geq p(p+q) & \quad \, \forall \emptyset \neq R \subsetneq V\, .  
\end{flalign}
This is also the starting point for Chekuri and Jain \cite{CJ25}. 
However, to obtain a valid IP, they additionally include the following ``robustness'' constraint:
\begin{flalign} \label{eq:robust}
\quad x(\delta(R) \sm F') \geq p & \quad \forall \, \emptyset\neq R\subsetneq V, \, \forall \, F' \subseteq \delta(R) \cap \unsafe \text{ with } |F'| \leq q\, .
\end{flalign}
Chekuri and Jain also show that the separation oracle for the associated LP can be implemented efficiently (see Lemma~1 in \cite{CJ25}).  

Our approach, instead of~\eqref{eq:robust}, is to include knapsack cover inequalities arising from \eqref{eq:basiccapineq}.
As mentioned before, this insight was inspired by the work of \cite{BCKS25} on $\oneqfgc$. 
We also highlight that Chakrabarty, Chekuri, Khanna, and Korula~\cite{CCKK15} used knapsack cover inequalities in the context of designing approximation algorithms for $\capkecss$. 
The natural IP formulation for $\capkecss$ has a cut-capacity constraint $\sum_{e \in \delta(R)} u_e x_e \geq k$ for each $\emptyset \neq R \subsetneq V$. 
It is not difficult to see that the associated LP may have an integrality gap of $\Omega(k)$. 
Chakrabarty et al. observed that including knapsack cover inequalities strengthens the associated LP. 

Before formulating the knapsack cover inequalities in \Cref{capKCineq} below, we provide some motivation. 
For this informal discussion, let us fix an $\emptyset \neq R \subsetneq V$ and focus solely on the (cut) edges in $\delta(R)$. 
We assume that $q$ is a large integer multiple of $p\ge 2$. 
Consider the following two solutions $F_1$ and $F_2$:  
$F_1$ has 1 safe edge and $p+q-1-q/p$ unsafe edges, while $F_2$ has $p-1$ safe edges and $1+q/p$ unsafe edges from $\delta(R)$.
Clearly, both $F_1$ and $F_2$ satisfy the basic cut-capacity constraint \eqref{eq:basiccapineq} for $R$, while both are infeasible for $\pqfgc$. 
We now show how a slightly modified version of \eqref{eq:basiccapineq} can detect the infeasibility of $F_1$. 
Let $J_1 := \delta_{F_1}(R) \cap \safe$ consist of the unique safe edge in $F_1$. 
Since $J_1$ already contributes a safe edge to $\delta(R)$, the residual requirement (w.r.t. $\pqfgc$ feasibility) from $F_1 \sm J_1$ is to have $p-1$ safe edges or $p+q-1$ total edges. 
Thus, we could rewrite \eqref{eq:basiccapineq} by giving a capacity of $p+q-1$ to safe edges and $p-1$ to unsafe edges and requiring that a solution has total capacity at least $(p-1)(p+q-1)$ on this cut. 
Since $(p-1)(p+q-1-q/p) < (p-1)(p+q-1)$, $F_1$ violates this modified version of \eqref{eq:basiccapineq}. 
Similarly, let $J_2 := \delta_{F_2}(R) \cap \unsafe$ consist of exactly $1+q/p$ unsafe edges in $F_2$. 
Since $J_2$ already contributes $1+q/p$ edges to $\delta(R)$, the residual requirement of $F_2 \sm J_2$ is to have $p$ safe edges or $p+q-1-q/p$ total edges. 
Again, as $(p-1)(p+q-1-q/p) < p(p+q-1-q/p)$ holds, $F_2$ violates a different version of \eqref{eq:basiccapineq}. 
Thus, adjusting the capacities of safe and unsafe edges in a cut depending on the residual requirement can strengthen the formulation. 
In fact, we show that these adjustments are sufficient to obtain a valid IP formulation. 
Recall that $z^+$ is shorthand for $\max(0, z)$ for any $z \in \R$. 

\begin{lemma} \label{capKCineq}
An edge set $F \subseteq E$ is a feasible solution for $\pqfgc$ if and only if for every $\emptyset \neq R \subsetneq V$ and every partition $\delta_F(R) = J \cup K$, the following inequality holds: 
\begin{equation} \label{eq:capKCineq}
(p - |J \cap \safe|)^+ \cdot |K| + (q - |J \cap \unsafe|)^+ \cdot |K \cap \safe| \geq (p - |J \cap \safe|)^+ \cdot (p + q - |J|)^+.
\end{equation}
\end{lemma}

\begin{proof}
For the implication, suppose $F$ is a feasible solution, $\emptyset \neq R \subsetneq V$ is a nontrivial set of vertices, and $J$ is a subset of edges in $\delta_F(R)$. 
Since the left-hand side (LHS) of \eqref{eq:capKCineq} is always nonnegative, it suffices to focus on the cases where $|J \cap \safe| < p$ and $|J| < p + q$ hold. 
By the feasibility of $F$, either $|\delta_F(R)| \geq p+q$ or $|\delta_F(R) \cap \safe| \geq p$ must hold (see \Cref{cutchar-prop}). 
Since $K = \delta_F(R) \sm J$, we have that either $|K| \geq p+q-|J|$ or $|K \cap \safe| \geq p - |J \cap \safe|$ holds. 
The inequality is trivial in the former case and holds in the latter case because $(p - |J \cap \safe|) + (q - |J \cap \unsafe|)^+ \geq p + q - |J|$.  
Hence, \eqref{eq:capKCineq} holds for every choice of $R$ and $J$ when $F$ is feasible. 

For the converse, suppose $F$ is an infeasible solution. 
By \Cref{cutchar-prop}, there is an $\emptyset \neq R \subsetneq V$ such that both $|\delta_F(R)| < p+q$ and $|\delta_F(R) \cap \safe| < p$ hold. 
Taking $J = \delta_F(R)$ and $K = \emptyset$, we see that the LHS of \eqref{eq:capKCineq} is $0$, while the right-hand side (RHS) is positive. 
This gives a violation in \eqref{eq:capKCineq}, as desired. \customqed  
\end{proof}

Our $O(\log n)$-approximation algorithm for $\pqfgc$ is based on independent randomized rounding of an optimal LP solution. 
The output of our algorithm is feasible with constant probability. 
The following result, borrowed from \cite{BCHI24}, will be useful in testing the feasibility of $F$. 

\begin{lemma}[Proposition~5.1 in \cite{BCHI24}] \label{feascheck}
For any set of edges $F \subseteq E$, we can efficiently check the feasibility of $F$ for the given instance of $\pqfgc$. 
\end{lemma}

Consider a capacitated graph $H = (V,F,u)$, where $u : F \to \Q_{>0}$ is a capacity function on the edges of $H$. 
Let 
\[
\lambda(H) := \min_{\emptyset \neq R \subsetneq V} \sum_{e \in F \cap \delta(R)} u_e
\]
denote the minimum cut of $H$. 
Note that $\lambda(H) > 0$ if and only if $(V,F)$ is connected. 
For any $\alpha \geq 1$, we say that $\emptyset \neq R \subsetneq V$ is an $\alpha$-approximate min-cut of $H$ whenever $\sum_{e \in F \cap \delta(R)} u_e \leq \alpha \cdot \lambda(H)$ holds. 
In a seminal work, Karger \cite{Karger93} showed that the number of $\alpha$-approximate min-cuts is $O(n^{2\alpha})$; the same work also gives a randomized polynomial-time algorithm to enumerate all $\alpha$-approximate min-cuts, with high probability, in $n^{O(\alpha)}$ time. 
Subsequently, Nagamochi, Nishimura, and Ibaraki \cite{NagamochiNI97} gave a deterministic enumeration algorithm in $n^{O(\alpha)}$ time. 
Very recently, Beideman, Chandrasekaran, and Wang \cite{BeidemanCW23} showed that any $\alpha$-approximate min-cut (of a connected graph) is the unique minimum $(S,T)$-terminal cut for some subsets $S$ and $T$ of vertices each of size at most $\lfloor 2 \alpha \rfloor + 1$. 
This elegant characterization lends itself to a simple deterministic enumeration of all approximate min-cuts.  
We paraphrase their main theorem for our purposes. 

\begin{theorem}[Follows from Theorem~1.1 and Remark~1.1 in \cite{BeidemanCW23}] \label{nearmincuts}
For any capacitated graph $H$ with nonzero $\lambda(H)$ and for any $\alpha \geq 1$, all $\alpha$-approximate min-cuts of $H$ can be enumerated in deterministic $n^{O(\alpha)}$ time. 
\end{theorem}

\Cref{nearmincuts} with $\alpha = O(1)$ is applied several times in our proofs. 
First, in \Cref{feascheck}, to check the feasibility of a candidate $\pqfgc$ solution by relating it to an instance of the $\capkecss$ problem; see \cite{BCHI24} for more details.
Second, in our separation oracle for checking if a candidate fractional solution is feasible for the LP relaxation (see \Cref{seporacle}).
Third, in the proof of the rounding result (\Cref{rounding}), where we use a union-bound argument over all cuts to show that the output of our algorithm is feasible with constant probability. 
\Cref{nearmincuts} with $\alpha = O(\ell)$ is used in Section~\ref{extension} while deriving results for the more general $\vfgc$ problem. 

\subsection{Other Related Work} 

The notion of introducing nonuniformity in network design via safe and unsafe edges is due to Adjiashvili \cite{adjiashvili13flexpaths}, where they initially considered flexible versions of the edge-disjoint $(s,t)$-paths problem. 
In a follow-up work, Adjiashvili, Stiller, and Zenklusen \cite{ASZ15} introduced a somewhat more general model called bulk-robust combinatorial optimization to capture highly correlated failure scenarios.  
Here, the input comes with an explicit list of $m$ failure scenarios $F_1,\ldots,F_m$, where a scenario specifies the resources that will fail simultaneously. 
The goal is to find a minimum-cost solution that remains feasible no matter which failure scenario occurs. 
A disadvantage of bulk-robust versions of connectivity problems is that failure scenarios have to be explicitly listed, while in the $\fgc$ model the failure
scenarios are implicitly encoded. 
However, $\pqfgc$ with constant $q$ can still be cast as an instance of the bulk-robust $p$-$\ecss$ problem with $m = \binom{|\unsafe|}{q} = |E|^{O(q)}$ failure scenarios. 
In terms of approximability, the two models have a stark separation for the vanilla $1$-connectivity problem. 
Recall that Bansal et al. \cite{BCKS25} gave an $8$-approximation algorithm for $\oneqfgc$ that holds for any $q$. 
However, the bulk-robust version of connectivity is set-cover hard, thereby ruling out sub-logarithmic approximations unless $\mathrm{P} = \mathrm{NP}$.  
See the proof of Proposition~1 in Section~2.1 of \cite{ASZ15} for more details. 

In \cite{CJ25}, Chekuri and Jain investigated local-connectivity versions of the $\fgc$ model, which they refer to as $(p,q)$-Flex-ST. 
Here, the input consists of source-sink pairs $\{ (s_i,t_i) \}_i$, each with prescribed edge-connectivity $p_i$ and robustness $q_i$ satisfying $p_i \leq p$ and $q_i \leq q$, and the goal is to find a minimum-cost $F \subseteq E$ such that for any $i$, the subgraph $(V,F)$ contains $p_i$ edge-disjoint $(s_i,t_i)$-paths regardless of which $q_i$ unsafe edges fail. 
In sharp contrast with the global connectivity version, even deciding the feasibility of a solution for $(p,q)$-Flex-ST is NP-complete, when $p$ and $q$ are part of the input; see Theorem~2 in \cite{HommelsheimLMZ25} for more details.  
In \cite{CJ25}, the authors give constant-factor approximation algorithms for $(p,q)$-Flex-ST when either $p$ or $q$ is sufficiently small, and a $\poly(p,q,\log n)$-approximation for the general case. 
We remark that the running time of these algorithms has an exponential dependence on $p$ and/or $q$, unlike the current approximation algorithms for various $\pqfgc$ regimes which all run in polynomial time. 

The FGC model has also been studied in the unweighted setting, i.e., when all edges have unit cost. 
Let $\alpha_k$ denote the current best approximation ratio for the unweighted $\kecss$ problem. 
Boyd et al. \cite{BCHI24} gave a $4\alpha_2/(2\alpha_2+1)$-approximation for unweighted $\oneonefgc$.
Using the result of \cite{HommelsheimLL26}, this implies an approximation ratio of $<1.4286$. 
Recently, Hyatt-Denesik, Jabal-Ameli, and Sanit\`{a} \cite{hyatt24} tightened the analysis of \cite{BCHI24} to obtain an approximation ratio of $10/7$. 
They also gave a $1+O(1/\sqrt{q})$-approximation algorithm for unweighted $\oneqfgc$.
Nutov \cite{Nutov25} gave an $(\alpha_p + 2q/p)$-approximation algorithm for unweighted $\pqfgc$. 

In their work, Hyatt-Denesik et al. \cite{hyatt24} introduced and studied the vertex connectivity version of FGC. 
Here, vertices are either safe or unsafe, and the goal is to find a minimum-size spanning subgraph that is connected and has the property that no unsafe vertex is a cut-vertex in this subgraph. 


The FGC model on directed graphs has received relatively little attention, largely due to the polylogarithmic hardness-of-approximation barriers that arise in directed survivable network design problems. 
Adjiashvili, Hommelsheim, M\"uhlenthaler, and Schaudt~\cite{AHMS22} study the $(s,t)$-connectivity problem in the $(1,k)$- and $(k,1)$-regimes, which they term \emph{Fault Tolerant Path} and \emph{Fault Tolerant Flow}, respectively.  
They give exact algorithms for the $(1,1)$-Flex-ST problem on general digraphs and for the $(1,k)$-Flex-ST problem on acyclic digraphs, and obtain a $k$-approximation for general $(1,k)$-Flex-ST instances. 
Complementing these results, they show that $(1,k)$-Flex-ST (respectively, $(k,1)$-Flex-ST) is at least as hard as Directed Steiner Tree (respectively, Directed Steiner Forest), implying polylogarithmic~\cite{HalperinK03} (respectively, almost polynomial~\cite{DodisK99}) hardness of approximation. 
These hardness results apply when $k$ is part of the input. 

Recently, Hommelsheim, Liu, Megow, and Zhang~\cite{HommelsheimLMZ25} investigated a ``connectivity preservation'' variant of the $\fgc$ model where each edge in the input graph is initially unsafe and there is an associated cost for upgrading this edge to a safe edge. 
They give several exact and approximation algorithms in this setting, along with some hardness-of-approximation results. 
We remark that the two models are incomparable. 

\section{Integer Programming Formulation for \boldmath \texorpdfstring{$\pqfgc$}{(p,q)-FGC}} \label{formulation}

As mentioned in the introduction, a key contribution in this work is a new IP formulation for $\pqfgc$ based on \Cref{capKCineq}. 
For each edge $e \in E$, we have a Boolean variable $x_e$ that denotes the inclusion of $e$ in a feasible solution $F$, i.e., $x_e = 1$ if and only if $e \in F$. 
Linearizing \eqref{eq:capKCineq} yields the following valid IP. 
By \Cref{capKCineq}, a 0-1 solution is feasible if and only if it corresponds to a feasible set $F$.

\noindent
\begin{minipage}[t]{\textwidth} 
\vspace{-10pt}
\begin{flalign}
\text{minimize } \, & \quad \sum_{e \in E} c_e x_e & &  \label{eq:lpobj} \\
\text{subject to:} & \quad (p-|J \cap \safe|)^+ \cdot x(K) \, + \, (q-|J \cap \unsafe|)^+ \cdot x(K \cap \safe) & & \notag \\ 
& \qquad \qquad \qquad \geq \, (p-|J \cap \safe|)^+ \cdot (p + q - |J|)^+ & & \notag \\ 
& \qquad \qquad \qquad \qquad \qquad \quad \, \, \forall \, \emptyset \neq R \subsetneq V, \, \forall \text{ partition } \delta(R) = J \cup K \label{eq:lpcovering} \\
& \quad 0 \leq x_e \leq 1, \quad x_e \in \Z \quad \, \forall e \in E  \label{eq:lpbox}
\end{flalign}
\end{minipage}

\medskip

Our linear program (LP) for $\pqfgc$ is obtained by relaxing the integrality condition on the $x_e$'s, so that they may take any value in $[0,1]$. 
Let $\lpopt$ denote the optimum value of this LP.  
\begin{lemma} \label{lplb}
We have $\lpopt \leq \opt$. 
\end{lemma}
 
Several remarks are in order. 
Although the number of constraints in the LP is exponential in $|E|$, we will show that it admits an efficient separation oracle. 
Here, we borrow ideas from Chekuri and Jain's work \cite{CJ25} (see Lemma~1).  
For any choice of $\emptyset \neq R \subsetneq V$ and the fixed choice of $J = \emptyset$, constraint~\eqref{eq:lpcovering} reduces to the covering inequality: ${p \cdot x(\delta(R)) + q \cdot x(\delta(R) \cap \safe) \geq p (p+q)}$.
These constraints are much simpler to work with and will be crucial in designing an efficient separation oracle; see \Cref{seporacle}. 

For an $x \in [0,1]^E$, we define the capacity function $w_x : E \to \Rp$ on the edges of $G$ as follows: $w_x(e) := (p+q \cdot \boldsymbol{1}_{e \in \safe}) x_e$ for any $e \in E$. 
For convenience, we refer to the graph $G$ equipped with the capacity function $w_x$ as the capacitated graph $H_x$. 
The following lemma gives two conditions, one necessary and the other sufficient, for $x$ to be a feasible LP solution.  

\begin{lemma} \label{feasmincut}
Let $x \in [0,1]^E$ and $H_x = (G,w_x)$ be the capacitated graph derived from $G$ and $x$. 
We have the following:
\begin{enumerate}[(a)]
    \item If $w_x(\delta(R)) < p(p+q)$ for some $\emptyset \neq R \subsetneq V$, then $x$ violates  constraint~\eqref{eq:lpcovering} for this choice of $R$ and $J=\emptyset$. 
    Therefore, if the minimum cut in $H_x$ is less than $p(p+q)$, then $x$ is infeasible for the LP.
    
    \item If $w_x(\delta(R)) \geq 2p(p+q)$ for some $\emptyset \neq R \subsetneq V$, then $x$ satisfies constraint~\eqref{eq:lpcovering} for this choice of $R$ and for every choice of $J \subseteq \delta(R)$. 
    Therefore, $x$ is feasible for the LP if the minimum cut in $H_x$ is at least $2p(p+q)$.    
\end{enumerate}
\end{lemma}

\begin{proof}
For the first part, observe that when $J = \emptyset$ the LHS of \eqref{eq:lpcovering} is exactly $w_x(\delta(R))$ and the RHS is $p(p+q)$. 
So we get a violated inequality if $w_x(\delta(R)) < p(p+q)$ holds. 

For the second part, suppose $R$ is such that $w_x(\delta(R)) \geq 2p(p+q)$ holds and $J \subseteq \delta(R)$ is an arbitrary subset of edges in $\delta(R)$.   
Clearly, constraint~\eqref{eq:lpcovering} is satisfied when $|J \cap \safe| \geq p$ or $|J| \geq p+q$, so we assume otherwise. 
Thus, the RHS of the constraint is $(p-|J \cap \safe|) \cdot (p+q-|J|)$. 
By definition: 
\[
w_x(\delta(R)) = (p+q) \cdot x(\delta(R) \cap \safe) + p \cdot x(\delta(R) \cap \unsafe).
\]
Since $w_x(\delta(R)) \geq 2p(p+q)$, either $x(\delta(R) \cap \safe) \geq p$ or $x(\delta(R) \cap \unsafe) \geq p+q$ holds.  
In the former case, we have: 
\begin{flalign*}
\text{LHS of \eqref{eq:lpcovering}} & = (p-|J \cap \safe|)^+ \cdot x(\delta(R) \sm J) + (q-|J \cap \unsafe|)^+ \cdot x((\delta(R) \sm J) \cap \safe) \\
& \geq (p+q-|J|) \cdot x((\delta(R) \sm J) \cap \safe) \\
& \geq (p+q-|J|) \cdot \bigl( x(\delta(R) \cap \safe) - |J \cap \safe| \bigr) \tag{$x \in [0,1]^E$} \\
& \geq (p-|J \cap \safe|) \cdot (p + q - |J|) = \text{RHS of \eqref{eq:lpcovering}} \tag*{}
\end{flalign*}
In the latter case, we have:
\begin{flalign*}
\text{LHS} \quad & \geq (p-|J \cap \safe|) \cdot x(\delta(R) \sm J)  \\
& \geq (p-|J \cap \safe|) \cdot \bigl( x(\delta(R)) - |J| \bigr)  \tag{$x \in [0,1]^E$} \\
& \geq (p-|J \cap \safe|) (p + q - |J|) \quad = \text{RHS} \tag*{} 
\end{flalign*}  
Note that the final equality in both chains of inequalities is because of our assumption that $|J \cap \safe| < p$ and $|J| < p+q$. \customqed
\end{proof}

The next lemma is based on a well-known trick that is used to check covering inequalities en masse.  

\begin{lemma} \label{choiceofJ}
Let $x \in [0,1]^E$ and $\emptyset \neq R \subsetneq V$ be a nontrivial set of vertices. 
Let $L_s$ and $L_u$ be ordered lists of safe edges (respectively, unsafe) in $\delta(R)$ in nonincreasing order of their $x_e$ values, with ties broken arbitrarily. 
For any nonnegative integers $a \leq \min(p-1,|L_s|)$ and $b \leq \min(p+q-a,|L_u|)$, let $J_{a,b}$ denote the edge set consisting of the first $a$ edges in $L_s$ and the first $b$ edges in $L_u$. 
If $x$ satisfies constraint~\eqref{eq:lpcovering} for every choice of $J = J_{a,b}$, then $x$ satisfies the same constraint for every choice of $J \subseteq \delta(R)$. 
\end{lemma}
\begin{proof}
By the form of constraint~\eqref{eq:lpcovering}, it is clear that the numerical quantities that are relevant to us are the following: 
\[
\alpha_1 := (p-|J \cap \safe|)^+, \alpha_2 := (q-|J \cap \unsafe|)^+, \text{ and } \alpha_3 := (p-|J \cap \safe|)^+ \cdot (p+q-|J|)^+.
\]
Since the values of $a := |J \cap \safe|$ and $b := |J \cap \unsafe|$ completely determine the values of $\alpha_i$s, we can check constraint~\eqref{eq:lpcovering} by grouping the possible choices of $J$ for each $a \leq \min(p-1,|\delta(R) \cap \safe|)$ and $b \leq \min(p+q-1,|\delta(R) \cap \unsafe|)$; note that if $\alpha_3 = 0$, then the constraint is trivially satisfied.  
Fix one such choice of $a$ and $b$.  
By grouping variables according to whether they are safe or unsafe, constraint~\eqref{eq:lpcovering} can be re-written as: 
\[
(\alpha_1 + \alpha_2) \cdot \sum_{e \in \delta_\safe(R) \sm J} x_e + \alpha_1 \cdot  \sum_{e \in \delta_\unsafe(R) \sm J} x_e \geq \alpha_3.
\]
Conditioned on $|J \cap \safe| = a$ and $|J \cap \unsafe| = b$, the set $J_{a,b}$ minimizes the LHS of the above inequality. 
So, if constraint~\eqref{eq:lpcovering} is satisfied by $J_{a,b}$, then it is satisfied by all such $J$. 
It follows that if constraint~\eqref{eq:lpcovering} is satisfied by $J_{a,b}$ for all choices of $a$ and $b$, then it is satisfied for every choice of $J \subseteq \delta(R)$. \customqed
\end{proof}

The above two lemmas imply a natural strategy for devising a separation oracle for our LP. 

\begin{lemma} \label{seporacle}
Let $x \in [0,1]^E$. 
In $\poly(n,p,q)$ time, we can determine whether $x$ satisfies constraint~\eqref{eq:lpcovering} for every choice of a $\emptyset \neq R \subsetneq V$ and a set of edges $J \subseteq \delta(R)$, or find a violated inequality.
\end{lemma}
\begin{proof}
Let $H_x = (G,w_x)$ denote the capacitated graph derived from $G$ and $x$. 
Using standard tools, we can efficiently compute a minimum cut in $H_x$.  
Let $R^*$ denote the set of vertices corresponding to a minimum cut, and $\lambda := w_x(\delta(R^*))$ denote its capacity.  
If $\lambda < p(p+q)$, then we get a violated inequality for $(R = R^*,J = \emptyset)$ by using the first part of \Cref{feasmincut}. 

Now suppose $\lambda \geq p(p+q)$. 
By the second part of \Cref{feasmincut}, any $\emptyset \neq R \subsetneq V$ with $w_x(\delta(R)) \geq 2p(p+q)$ cannot give a violated inequality for any choice of $J \subseteq \delta(R)$. 
Thus, it suffices to focus only on those $R$ with $w_x(\delta(R)) < 2p(p+q)$. 
Define 
\[
\Rs := \{ \emptyset \neq R  \subsetneq V : w_x(\delta(R)) < 2p(p+q) \}.
\]
Note that $|\Rs| = O(n^4)$ (\Cref{nearmincuts}) by our assumption that $\lambda \geq p(p+q)$. 
In fact, using \cite{NagamochiNI97}, we can efficiently enumerate $\Rs$ in $\poly(n)$ time. 
Now fix some $R \in \Rs$. 
Let $L_s$ and $L_u$ be the ordered lists of the safe (respectively, unsafe) edges in $\delta(R)$ in nonincreasing order of their $x_e$ values, with ties broken arbitrarily. 
By \Cref{choiceofJ}, if there is a violation in \eqref{eq:lpcovering} for some $J$, then there is a violation for some suitable $J_{a,b}$. 
Thus, we either find a violated inequality for this choice of $R$, if one exists, or we have a certificate that there are no violations for any choice of $J$. \customqed
\end{proof}

In summary, $x$ is feasible if and only if all the following conditions hold: (i)~$x \in [0,1]^E$; (ii)~the minimum cut in $H_x$ is at least $p(p+q)$; and (iii)~constraint~\eqref{eq:lpcovering} holds for every $(R,J)$, where $R$ has $w_x(\delta(R)) < 2p(p+q)$ and $J$ consists of up to $p-1$ safe edges and up to $p+q-1$ unsafe edges with the highest value of $x_e$.  
Since the feasible region has bounded facet complexity (and is, in particular, in the box $[0,1]^E$), and an efficient separation oracle is available, we can use the Ellipsoid method to obtain an optimal solution in polynomial time; see \cite[Chapter 6]{GLS81}.

\begin{theorem} \label{lpsolver}
Suppose that we are given a feasible instance of the $\pqfgc$ problem.  
We can efficiently compute a vector $x \in [0,1]^E$ that is feasible for the linear relaxation of the integer program given by \eqref{eq:lpobj}--\eqref{eq:lpbox} and has $c^\top x := \sum_{e \in E} c_e x_e$ at most \opt.  
\end{theorem}

\section{Independent Rounding Algorithm for \boldmath \texorpdfstring{$\pqfgc$}{(p,q)-FGC}} \label{indepround}

Our main result in this section is an algorithm that takes as input a fractional LP solution $x$ and that, with constant probability, outputs a random edge set $F$ that is an $O(\log n)$-approximate solution for the given instance of $\pqfgc$. 
Our algorithm is conceptually simple, relying on independent randomized rounding. 
Define $y \in [0,1]^E$ as follows: for each edge $e$, 
\[
y_e := \min(1,(100 \log n) x_e).
\]
Let $F$ be a random set of edges that is obtained by independently including each edge $e \in E$ in $F$ with probability $y_e$. 

\begin{theorem} \label{rounding} 
With probability at least $1/3$, $F$ is feasible and $c(F) \leq (200 \log n) \cdot c^\top x$. 
\end{theorem}

Running the algorithm with an optimal LP solution $x$ obtained from \Cref{lpsolver} gives us an $O(\log n)$ approximation algorithm for $\pqfgc$, proving our headline result \Cref{logapx}. 
We can efficiently check whether $F\subseteq E$ is a feasible solution for the given $\pqfgc$ instance (\Cref{feascheck}); this involves enumerating all near-minimum cuts, as in the proof of \Cref{seporacle}. 
Thus, the algorithm can be run repeatedly until a feasible solution with cost $O(\log n) \cdot c^\top x$ is obtained.  
This gives a Las Vegas $O(\log n)$-approximation algorithm with expected running time that is polynomial in the input encoding size. 

\medskip

We give a proof of \Cref{rounding} at the end of this section. 
Like in \cite{CCKK15,CJ25}, we use the well-known technique of applying a union-bound argument over all cuts in a graph. 
The following version of the Chernoff lower tail bound will be useful to us; for example, see \cite{MU-book}.

\begin{theorem}[Chernoff bound] \label{chernoff} 
Let $\{X_j\}_j$ be a finite collection of independent random variables taking values in $\{0,1\}$, and let $Z$ denote their sum. 
For any $\mu \leq \E[Z]$ and $\eta \in (0,1)$, we have: 
\[
\Pr[Z < (1-\eta)\mu] \leq \exp(-\mu \eta^2/2).
\]
In particular, by taking $\eta = 2/3$, we can derive $\Pr[Z < \mu/3] \leq \exp(-\mu/5)$.
\end{theorem}

Define $A := \{e \in E : y_e = 1\}$ as the set of edges (safe or unsafe) that are always included in $F$.  
By definition, for any $e \notin A$ we have $y_e = (100 \log n) x_e < 1$. 
For any $\emptyset \neq R \subsetneq V$, define $B_R$ as the indicator random variable for the \emph{bad} event $\{ |\delta_F(R) \cap \safe| < p \text{ and } |\delta_F(R)| < p+q \}$.  
Note that $B_R = 1$ whenever the cut $\delta(R)$ certifies the infeasibility of $F$.  
The following lemma is trivial. 

\begin{lemma} \label{feaslargevars}
If $|\delta_A(R) \cap \safe| \geq p$ or $|\delta_A(R)| \geq p+q$ holds, then $\Pr[B_R = 1] = 0$.    
\end{lemma}

Given the above lemma, it suffices to focus only on those $\emptyset \neq R \subsetneq V$ for which both $|\delta_A(R) \cap \safe| < p$ and $|\delta_A(R)| < p+q$ hold.  
Since $x$ is a feasible LP solution, it satisfies constraint~\eqref{eq:lpcovering} for this choice of $R$ and all choices of $J \subseteq \delta(R)$.  
For convenience, we restate constraint~\eqref{eq:lpcovering} below:
\begin{flalign} 
\bigl( (p-|J \cap \safe|) + (q-|J \cap \unsafe|)^+ \bigr) \cdot x(K \cap \safe) + (p-|J \cap \safe|) \cdot x(K \cap \unsafe) \notag \\ \geq (p-|J \cap \safe|) (p + q - |J|). \label{eq:xmassforsmallcuts}
\end{flalign}

As the edges in $A$ have been purchased fully and the LP solution $x$ satisfies all knapsack cover inequalities, it is natural to consider the partition $\delta(R) = J \cup K$ where $J := A \cap \delta(R)$. 
As none of the $A$-edges appear in $K$, we have $y_e < 1$ for every $e \in K$. 
Changing the variable in \eqref{eq:xmassforsmallcuts} from $x$ to $y$ gives:
\begin{flalign} 
\bigl( (p-|J \cap \safe|) + (q-|J \cap \unsafe|)^+ \bigr) \cdot y(K \cap \safe) + (p-|J \cap \safe|) \cdot y(K \cap \unsafe) \notag \\ \geq (100 \log n) (p-|J \cap \safe|) (p + q - |J|). \label{eq:meanforsmallcuts}
\end{flalign}

For an edge $e \in K$, let $Y_e$ denote the indicator random variable for the event $\{e \in F\}$. 
Note that $\E[Y_e] = y_e = (100 \log n) x_e$. 
Define the integer random variables $Z_s := \sum_{e \in K \cap \safe} Y_e$ and $Z_u := \sum_{e \in K \cap \unsafe} Y_e$ to be the number of safe and unsafe $K$-edges, respectively, that are included in $F$. 
The following lemma gives a crude upper bound on $\Pr[B_R = 1]$ for any $R$.  

\begin{lemma} \label{badsmallcut}
We have $\Pr[B_R = 1] \leq n^{-10}$. 
\end{lemma}
\begin{proof}
The lemma is trivial if $R$ satisfies the hypothesis of \Cref{feaslargevars}, so we assume otherwise. 
By the above discussion, \eqref{eq:meanforsmallcuts} implies that at least one of the following three cases occurs.  
\begin{enumerate}
\item Suppose $|J \cap \unsafe| < q$ and $y(K \cap \safe) \geq 50 \cdot \log n \cdot (p-|J \cap \safe|) =: \mu_1$ hold.
Observe that: 
\begin{align*}
\Pr[B_R = 1] & \leq \Pr[|\delta_F(R) \cap \safe| < p] & \\
& = \Pr[|F \cap K \cap \safe| < p - |J \cap \safe|] & & \\ 
& = \Pr[Z_s < p - |J \cap \safe|] \leq \Pr[Z_s < \mu_1 / 3] \tag{$\E[Z_s] = y(K \cap \safe)$} \\ 
& \leq \exp(-(10 \log n)(p-|J \cap \safe|)) \tag{\Cref{chernoff}} \\
& \leq n^{-10}. \tag{$p - |J \cap \safe| \geq 1$} 
\end{align*}

\item Suppose $|J \cap \unsafe| < q$ and $y(K \cap \unsafe) \geq (50 \log n) (p + q - |J|) =: \mu_2$ hold. 
We repeat the same calculation as above for the random variable $Z_u$, whose expectation $y(K \cap \unsafe)$ is at least $\mu_2$, to get:
\begin{align*}
\Pr[B_R = 1] & \leq \Pr[|\delta_F(R)| < p+q] = \Pr[|F \cap K| < p + q - |J|] & \\
& \leq \Pr[|F \cap K \cap \unsafe| < p + q - |J|] = \Pr[Z_u < p + q - |J|]  \\
& \leq \Pr[Z_u < \mu_2 / 3] \leq \exp(-(10 \log n) (p+q-|J|)) \leq n^{-10}. 
\end{align*}

\item Suppose $|J \cap \unsafe| \geq q$ and $y(K) \geq 100 \cdot \log n \cdot (p + q - |J|) =: \mu_3$ hold.
We repeat the same calculation for the random variable $Z_s + Z_u$, whose expectation $y(K)$ is at least $\mu_3$, to obtain $\Pr[B_R = 1] \leq \Pr[Z_s + Z_u < p + q - |J|] \leq n^{-20}$.
\end{enumerate}

Thus, the bad event associated with $B_R$ occurs with probability at most $n^{-10}$. \customqed
\end{proof}

We now derive a much stronger tail bound for cuts with large $w_x$ capacity. 

\begin{lemma} \label{badlargecut}
Suppose $R$ is such that $w_x(\delta(R)) \geq \tau p(p+q)$
for some $\tau \geq 4$. 
We have $\Pr[B_R = 1] \leq n^{-5\tau}$.  
\end{lemma}
\begin{proof}
We follow the same template as in the proof of \Cref{badsmallcut}. 
We first show that $x(K \cap \safe) \geq \tau p/4$ or $x(K \cap \unsafe) \geq \tau (p+q)/4$ holds and then use Chernoff tail bounds to establish the claim.   
The lemma is trivial if $R$ satisfies the hypothesis of \Cref{feaslargevars}, so we assume otherwise. 
By the definition of $w_x$, we have: 
\[
w_x(\delta(R)) = (p+q) \cdot x(\delta(R) \cap \safe) + p \cdot x(\delta(R) \cap \unsafe).
\]
Since $w_x(\delta(R)) \geq \tau p(p+q)$, either $x(\delta(R) \cap \safe) \geq \tau p/2$ or $x(\delta(R) \cap \unsafe) \geq \tau(p+q)/2$ holds. 
By our assumption on $R$, we have $x(J \cap \safe) \leq |J \cap \safe| < p$ and $x(J \cap \unsafe) \leq x(J) \leq |J| < p+q$ since $x \in [0,1]^E$.  
As $J$ and $K$ partition $\delta(R)$, either $x(K \cap \safe) \geq \tau p/4$ or $x(K \cap \unsafe) \geq \tau (p+q)/4$ holds. 
Here, we use $\tau \geq 4$. 
It remains to apply Chernoff bounds in each of the two cases. 
We recall that $y_e = (100 \log n) x_e < 1$ for each $e \in K$. 
\begin{enumerate}

\item Suppose $y(K \cap \safe) \geq (25 \tau \log n) p =: \mu_1$ holds. 
We have: 
~
\begin{flalign*}
\Pr[B_R = 1] & \leq \Pr[|\delta_F(R) \cap \safe| < p] \leq \Pr[|F \cap K \cap \safe| < p] & \\ 
& \leq \Pr[Z_s < \mu_1 / 3] \tag{$\E[Z_s] = y(K \cap \safe)$} \\ 
& \leq \exp(-(5 \tau \log n) p) \leq n^{-5\tau} \tag{\Cref{chernoff}}
\end{flalign*}

\item Suppose $y(K \cap \unsafe) \geq (25 \tau \log n) (p + q) =: \mu_2$ holds. 
Again, we have: 
\begin{flalign*}
\Pr[B_R = 1] & \leq \Pr[|\delta_F(R)| < p+q] \leq \Pr[|F \cap K \cap \unsafe| < p + q] & \\
& \leq \Pr[Z_u < \mu_2 / 3] \tag{$\E[Z_u] = y(K \cap \unsafe)$} \\ 
& \leq \exp(-(5 \tau \log n)(p+q)) \leq n^{-5\tau}. \tag{\Cref{chernoff}}
\end{flalign*}
\end{enumerate}
Thus, in both cases, the bad event $B_R$ occurs with probability at most $n^{-5\tau}$. \customqed
\end{proof}

We are now ready to prove the main result
of this section. 

\begin{proofof}{\Cref{rounding}}
First, observe that the expected cost of $F$ is at most $(100 \log n) \cdot c^\top x$ since each edge $e$ is included in $F$ with probability $y_e \leq (100 \log n) x_e$. 
By Markov's inequality, $c(F) \geq (200 \log n) \cdot c^\top x$ holds with probability at most $1/2$. 
We assume that this does not happen. 

Second, we use a union-bound argument to show that $F$ is infeasible with probability at most $1/6$. 
For $F$ to be infeasible, $B_R$ must be $1$ for some $\emptyset \neq R \subsetneq V$.  
We partition the nontrivial cuts in $H_x$ as $\C_{<4} \cup \C_4 \cup \C_5 \cup \ldots \cup \C_{|E|}$ where cuts in $\C_{<4}$ satisfy 
\[
w_x(\delta(R)) \in [p(p+q),4p(p+q))
\]
and for $\tau \in \Z_{\geq 4}$, cuts in $\C_\tau$ satisfy:  
\[
w_x(\delta(R)) \in [\tau p(p+q),(\tau+1)p(p+q)).
\]
By \Cref{nearmincuts} we have $|\C_{<4}| = O(n^8)$, where we use the fact that the minimum cut in $H_x$ is at least $p(p+q)$ (see the first part of \Cref{feasmincut}). 
In \Cref{badsmallcut} we showed that $\Pr[B_R = 1] \leq n^{-10}$ for any $R$.
Therefore: 
\[
\sum_{R \in \C_{<4}} \Pr[B_R = 1] = O(n^{-2}).
\]
Again, by \Cref{nearmincuts}, we have $|\C_{\tau}| \leq O(n^{2\tau+2})$ for any $\tau \in \Z_{\geq 4}$. 
In \Cref{badlargecut} we showed that $\Pr[B_R = 1] \leq n^{-5\tau}$ for any $R \in \C_\tau$. 
Thus, for any $\tau \geq 4$ we have: 
\[
\sum_{R \in \C_\tau} \Pr[B_R = 1] \leq O(n^{-3\tau + 2}) = O(n^{-2 \tau}).
\]
Overall, for sufficiently large $n$, we get: 
\begin{flalign*}
\sum_R \Pr[B_R = 1] & = \sum_{R \in \C_{<4}} \Pr[B_R = 1] + \sum_{\tau = 4}^{\infty} \Bigl( \sum_{R \in \C_{\tau}} \Pr[B_R = 1] \Bigr) \leq 1/6.
\end{flalign*} 
To summarize, $F$ is feasible for the given instance of $\pqfgc$ and has cost at most $(200 \log n) \cdot c^\top x$ with probability at least $1 - 1/2 - 1/6 = 1/3$. \customqed
\end{proofof}

\section{Extension of the \texorpdfstring{$\pqfgc$}{(p,q)-FGC} Model} \label{extension} 

We propose an extension of the $\pqfgc$ problem where the edges are partitioned into two or more tiers of safety. 
Formally, we are given an undirected multigraph $G = (V,E)$ with nonnegative edge costs $\{c_e\}_{e \in E}$, a partition $E = \cup_{i=1}^{\ell} E_i$ of the edge set into $\ell \in \Z_{\geq 1}$ tiers of decreasing safety, and an $\ell$-tuple $\vq = (q_1,\ldots,q_\ell) \in \Zp^{\ell}$ that encodes robustness parameters. 
The algorithmic goal is to find a minimum cost $F \subseteq E$ so that the spanning subgraph $(V,F \sm F')$ is connected for every choice of $F' \subseteq E$ satisfying $|F' \cap (E_1 \cup \ldots \cup E_i)| \leq q_1 + \cdots + q_i$ for all $i \in [\ell]$. 
We call this problem $\vfgc$, where the length of the robustness tuple $\vq$ implicitly encodes $\ell$, the number of safety tiers. 
Observe that the classical $\kecss$ problem corresponds to taking $\ell = 1$, $q_1 = k-1$, and $E_1 = E$. 
The $\pqfgc$ problem corresponds to taking $\ell = 2$, $q_1 = p-1$, $q_2 = q$, $E_1 = \safe$, and $E_2 = \unsafe$. 

We make a few assumptions about the instance. 
We will treat $\ell$ as a constant, but its influence on our main algorithm's approximation ratio or the running times is made explicit. 
Without loss of generality, we assume that $q_2,\ldots,q_\ell$ are all positive since otherwise a $0$-valued coordinate $q_i$ can be deleted by merging tier $E_i$ into tier $E_{i-1}$. 
For any $i \in [\ell]$, let $E_{\leq i}$ and $q_{\leq i}$ be shorthands for $\cup_{j=1}^{i} E_j$ and $\sum_{j=1}^{i} q_j$, respectively. 
Since we assume that $q_2,\ldots,q_\ell > 0$, $q_{\leq i}$ is strictly increasing over $i \in [\ell]$.  
Our motivation for studying this problem is to understand the ``uncapacitated'' version, so we will assume that $q_{\leq \ell} \leq n$. 
In particular, this assumption implies $|E| = \poly(n)$ as any optimal solution contains at most $1+q_{\leq i}$ parallel edges (between any two vertices) from each tier $E_i$ for $i \in [\ell]$.
Last, we assume that the instance is feasible. 
We will see shortly that the feasibility of $E$ can be checked in $n^{O(\ell)}$ time (\Cref{feascheck-ext}). 

Our main result in this section is an $O(\ell^2 \log n)$-approximation algorithm for $\vfgc$ (\Cref{logapx-ext}), which follows from a natural extension of the ideas that went into establishing a similar approximation guarantee for $\pqfgc$, which corresponds to the $\ell = 2$ case. 
We note that our main algorithm and many of the associated subroutines have a runtime with exponential dependence on $\ell$. 

We divide this section into three parts. 
In \Cref{characterization-ext}, we define the notion of a ``vulnerable partition'' of edges in a cut that will be useful in the characterization of feasible solutions for $\vfgc$. 
In \Cref{formulation-ext}, we give an integer programming formulation for $\vfgc$ such that the corresponding linear programming relaxation can be solved in $n^{O(\ell)}$ time. 
Finally, in \Cref{indepround-ext}, we show that independently rounding edge variables in a suitably scaled LP solution yields an $O(\ell^2 \log n)$-approximate solution for $\vfgc$.

\subsection{Characterization of \texorpdfstring{$\vfgc$ solutions}{multi-tier FGC solutions}} \label{characterization-ext}

For ease of presentation, we introduce the notion of vulnerable sets. 

\begin{definition} \label{def:vulnerable}
For any $\emptyset \neq R \subsetneq V$ and any $J \subset \delta(R)$, we say that $J$ is \emph{vulnerable} if $|J \cap E_{\leq i}| \leq q_{\leq i}$ holds for all $i \in [\ell]$. 
\end{definition}

In the context of $\pqfgc$, $J \subset \delta(R)$ is vulnerable if its cardinality is less than $p+q$ and it contains fewer than $p$ safe edges. 
For any $\emptyset \neq R \subsetneq V$, $J \subset \delta(R)$, and $K := \delta(R) \sm J$, we call the ordered pair $(R,J)$ (or the ordered partition $(J,K)$) vulnerable if $J$ is vulnerable. 
We often drop $K$ (or $R$) while referring to a vulnerable pair (or partition, respectively) because the choice is clear from the context. 
Note that for every $\emptyset \neq R \subsetneq V$, the trivial choice of $J = \emptyset$ (and $J = \delta(R)$) corresponds to an edge set that is always vulnerable (respectively, never vulnerable).  

\begin{proposition} \label{cutchar-ext}
An edge set $F \subseteq E$ is feasible for $\vfgc$ if and only if $\delta_F(R)$ is not vulnerable for any $\emptyset \neq R \subsetneq V$.  
\end{proposition}

\begin{proof}
For the implication, suppose $F$ is feasible and $\delta_F(R)$ is vulnerable for some $R$. 
By \Cref{def:vulnerable} and the feasibility of $F$, we must have that the spanning subgraph $(V,F \sm \delta_F(R))$ is connected. 
This leads to a contradiction, so no such $R$ exists. 

For the converse, suppose $F$ is such that $\delta_F(R)$ is not vulnerable for any $R$. 
By \Cref{def:vulnerable}, for every such $R$, there is a tier $i = i(R) \in [\ell]$ for which $|\delta_F(R) \cap E_{\leq i}| \geq 1 + q_{\leq i}$ holds.  
For any $F' \subseteq E$ that satisfies $|F' \cap E_{\leq i}| \leq q_{\leq i}$ for all $i$ and any $\emptyset \neq R \subsetneq V$, we have that $\delta_{F \sm F'}(R)$ is not empty since it contains the nonempty set $(\delta_F(R) \sm F') \cap E_{\leq i(R)}$. 
Therefore, $(V,F \sm F')$ is connected. 
So $F$ is feasible. \customqed
\end{proof}

Recall that the feasibility of an edge set $F$ for a given instance of $\pqfgc$ can be efficiently checked (\Cref{feascheck}). 
A natural extension of this result gives an algorithm for checking feasibility w.r.t. the given $\vfgc$ instance, although using $n^{O(\ell)}$ time. 

\begin{lemma} \label{feascheck-ext}
For any $F \subseteq E$, the feasibility of $F$ for the given instance of $\vfgc$ can be checked in $n^{O(\ell)}$ time. 
\end{lemma}
\begin{proof}
We construct an auxiliary capacitated graph $H = (V,F,u)$ where $u : F \to \Q_{>0}$ is a capacity function on the edges of $H$. 
The idea here is that testing all $\ell$-approximate minimum cuts in $H$ is sufficient to decide the feasibility of $F$.

For any tier $i \in [\ell]$ and any tier-$i$ edge $e \in E_i$, we define $u(e) := (1+q_{\leq i})^{-1}$.
Let $\lambda(F) := \min\{ u(\delta(R) \cap F) : \emptyset \neq R \subsetneq V \}$ denote the value of a minimum cut in $H$. 
The following necessary condition can be checked efficiently. 

\begin{claim} \label{infeasmincut-ext-clm}
If $\lambda(F) < 1$, then $F$ is infeasible. 
\end{claim}
\begin{proof}
Suppose, for the sake of contradiction, that $F$ is feasible. 
By the implication in \Cref{cutchar-ext}, for any $\emptyset \neq R \subsetneq V$ there is a tier $i = i(R)$ such that $|\delta_F(R) \cap E_{\leq i}| \geq 1+q_{\leq i}$ holds. 
Since every edge in $E_{\leq i}$ has capacity at least $1/(1+q_{\leq i})$, we get:
\[
u(\delta(R) \cap F) \geq |\delta_F(R) \cap E_{\leq i}|/(1+q_{\leq i}) \geq 1.
\]
So $\lambda(F) \geq 1$, as desired. 
\end{proof}

Now suppose $\lambda \geq 1$. 
Define: 
\[
\Rs := \{ \emptyset \neq R \subsetneq V : u(\delta(R) \cap F) \leq \ell \}
\]
to be the family of all near-minimum cuts in $H$ with capacity at most $\ell$.  
Note that $\Rs$ is a subfamily of all $\ell$-approximate min-cuts in $H$, so they can be enumerated in $n^{O(\ell)}$ time (recall \Cref{nearmincuts}). 
Our next claim gives a sufficient condition for the feasibility of $F$.

\begin{claim} \label{infeascert-ext}
If $F \subseteq E$ is infeasible, then $\delta_F(R)$ is vulnerable for some $R \in \Rs$. 
\end{claim}
\begin{proof}
By the inverse of \Cref{cutchar-ext}, $\delta_F(R)$ is vulnerable for some $\emptyset \neq R \subsetneq V$. 
We have:
\begin{align*}
u(\delta(R) \cap F) = \sum_{i=1}^{\ell} \Bigl( \frac{1}{1+q_{\leq i}} \cdot |\delta_F(R) \cap E_{i}| \Bigr) \leq \sum_{i=1}^{\ell} \frac{q_{\leq i}}{1+q_{\leq i}}  \leq \ell, 
\end{align*}
where we use \Cref{def:vulnerable} in the first inequality. 
So $R \in \Rs$, and we are done. 
\end{proof}

In summary, $F \subseteq E$ is feasible if and only if (a)~the capacity of a minimum cut in $H$ is at least $1$; and (b)~for every $\ell$-approximate minimum cut $R$ (in $H$), $\delta_F(R)$ is not vulnerable. 
The overall check can be performed in $n^{O(\ell)}$ time. \customqed
\end{proof}

Our next result is an extension of \Cref{capKCineq}, where we give an exact linear characterization of feasible solutions for the $\vfgc$ problem. 
The following notion of a residual capacity function will be relevant to us. 
For every $\emptyset \neq R \subsetneq V$ and every vulnerable partition $(J,K)$ of $\delta(R)$, we define a residual capacity function $u(\cdot;R,J)$ on $K$. 
For any tier $i \in [\ell]$ and any tier-$i$ edge $e \in K \cap E_i$ we define:
\begin{equation} \label{eq:rescapvecfgc}
u(e;R,J) := \max_{j \in [\ell] : j \geq i} (1 + q_{\leq j} - |J \cap E_{\leq j}| )^{-1}. 
\end{equation}
Since residual capacities are uniform within a tier $i \in [\ell]$, we overload $u(i;R,J)$ notation to refer to this common capacity value and define $u(\ell+1;R,J) := 0$ for convenience. 

\begin{claim} 
For any vulnerable pair $(R,J)$, the values $u(i;R,J)$ are finite and nonincreasing over $i \in [\ell]$. 
\end{claim}
\begin{proof}
The monotonicity follows from the recursive definition of $u(i;R,J)$. 
For any tier $i \in [\ell]$, we have:  
\begin{equation} \label{eq:recursiverescap}
u(i;R,J) = \max\bigl\{ u(i+1;R,J),(1+q_{\leq i}-|J \cap E_{\leq i}|)^{-1} \bigr\}.
\end{equation}
The residual capacities are finite because $1+q_{\leq i}-|J \cap E_{\leq i}| > 0$ for all $i$ by \Cref{def:vulnerable}. 
This finishes the proof. 
\end{proof}

The intuition behind our choice of residual capacities is that it suffices to purchase $1/u(i;R,J)$ many tier-$i$ edges from $K$ to ensure robustness w.r.t. the cut $\delta(R)$.  
The maximum is taken over all $j \geq i$ because we might be closer to attaining robustness w.r.t. a tier $j > i$ when $J$ already has many edges from $E_{\leq j}$ relative to the robustness requirement of $1+q_{\leq j}$; in the $\pqfgc$ setting, this was captured by the $(q-|J \cap \unsafe|)^+$ term in \eqref{eq:capKCineq}. \footnote{Residual capacities in $\pqfgc$ were not normalized w.r.t. the RHS of \eqref{eq:capKCineq}.}

\begin{lemma} \label{capKCineq-ext}
An edge set $F \subseteq E$ is feasible for $\vfgc$ if and only if for every $\emptyset \neq R \subsetneq V$ and every vulnerable partition $(J,K)$ of $\delta(R)$ the following inequality holds: 
\begin{equation} \label{eq:capKCineqvecfgc}
\sum_{e \in K \cap F} u(e;R,J) = \sum_{i \in [\ell]} u(i;R,J) \cdot |K \cap F \cap E_i| \geq 1.
\end{equation}
\end{lemma}
\begin{proof}
We first prove the inverse of the implication.
Suppose $F$ is infeasible.
By \Cref{cutchar-ext}, there is an $\emptyset \neq R \subsetneq V$ such that $J := \delta_F(R)$ is vulnerable. 
Note that $K := \delta(R) \sm J$ satisfies $K \cap F = \emptyset$.  
We get a violation in \eqref{eq:capKCineqvecfgc} as the (empty) sum in the LHS is $0$ while the RHS is $1$. 

Now suppose $F$ is feasible and $J \subset \delta(R)$ is vulnerable for some $\emptyset \neq R \subsetneq V$. 
By \Cref{cutchar-ext}, $\delta_F(R)$ is not vulnerable, so $|\delta_F(R) \cap E_{\leq i}| \geq 1 + q_{\leq i}$ for some tier $i \in [\ell]$. 
Observe that $K := \delta(R) \sm J$ satisfies $|K \cap F \cap E_{\leq i}| \geq 1 + q_{\leq i} - |J \cap E_{\leq i}|$. 
Since the residual capacities are nonincreasing over the tiers and $u(i;R,J) \geq (1 + q_{\leq i} - |J \cap E_{\leq i}|)^{-1}$ by definition, we get
\[
\sum_{e \in K \cap F} u(e;R,J) \geq |K \cap F \cap E_{\leq i}| \cdot u(i;R,J) \geq 1. 
\]
So \eqref{eq:capKCineqvecfgc} holds. \customqed 
\end{proof}

\subsection{Integer Programming Formulation for \texorpdfstring{$\vfgc$}{multi-tier FGC}} \label{formulation-ext}

\Cref{capKCineq-ext} motivates the following integer linear formulation for $\vfgc$. 
For each edge $e \in E$, we have a Boolean variable $x_e$ that models the inclusion of $e$ in some feasible solution $F$. 
Linearization of \eqref{eq:capKCineqvecfgc} yields the following valid IP. 
By \Cref{capKCineq-ext}, a 0-1 solution is feasible for the IP if and only if it corresponds to a feasible solution $F$.

\noindent
\begin{minipage}[t]{\textwidth} 
\begin{flalign}
\text{minimize } & \quad \sum_{e \in E} c_e x_e &  \label{eq:lpobj-ext} \\
\text{subject to:} & \quad \sum_{e \in K} u(e;R,J) \cdot x_e \geq 1 & \forall \text{ vulnerable partitions } \delta(R) = J \cup K \label{eq:lpcovering-ext} \\
& \quad 0 \leq x_e \leq 1, \, \, x_e \in \Z & \forall e \in E \label{eq:lpbox-ext}
\end{flalign}
\end{minipage}

\medskip

Our linear relaxation for $\vfgc$ is obtained by relaxing the integrality condition on the $x_e$'s, so that they may take any value in $[0,1]$. 
Let $\lpopt$ denote the optimum value of this LP.  
\begin{lemma} \label{lplb-ext}
We have $\lpopt \leq \opt$. 
\end{lemma}

For an $x \in [0,1]^E$, we define a capacity function $w_x : E \to \Rp$ on the edges of $G$ as follows: for any tier $i \in [\ell]$ and any tier-$i$ edge $e \in E_i$,  
\begin{equation} \label{eq:edgecap-ext}
w_x(e) := x_e \cdot (1+q_{\leq i})^{-1}.
\end{equation}
We refer to the graph $G$ equipped with the capacity function $w_x$ as the capacitated graph $H_x$. 
The following lemma gives two conditions, one necessary and the other sufficient, for $x$ to be a feasible LP solution.  

\begin{lemma} \label{feasmincut-ext}
Let $x \in [0,1]^E$ and $H_x = (G,w_x)$ be the capacitated graph derived from $G$ and $x$. 
We have the following:
\begin{enumerate}[(a)]
    \item If $w_x(\delta(R)) < 1$ for some $\emptyset \neq R \subsetneq V$, then $x$ violates  constraint~\eqref{eq:lpcovering-ext} for the trivial vulnerable partition $(\emptyset,\delta(R))$.  
    Therefore, if the minimum cut in $H_x$ is strictly less than $1$, then $x$ is infeasible for the $\vfgc$ LP. 
    
    \item If $w_x(\delta(R)) \geq \ell$ for some $\emptyset \neq R \subsetneq V$, then $x$ satisfies constraint~\eqref{eq:lpcovering-ext} for every vulnerable partition of $\delta(R)$. 
    Therefore, $x$ is feasible for the LP if the minimum cut in $H_x$ is at least $\ell$.    
\end{enumerate}
\end{lemma}

\begin{proof}
For the first part, observe that the LP inequality \eqref{eq:lpcovering-ext} for the trivial vulnerable partition $(\emptyset,\delta(R))$ can be rewritten as ${w_x(\delta(R)) \geq 1}$. 
This is because for any tier-$i$ edge $e \in \delta(R)$, we have: 
\[
u(e;R,\emptyset) \cdot x_e = u(i;R,\emptyset) \cdot x_e = (1+q_{\leq i})^{-1} \cdot x_e = w_x(e),
\]
where the equations rely on the assumption that $q_{\leq i}$ is strictly increasing in $i$. 
Therefore, we get a violated inequality whenever $w_x(\delta(R)) < 1$ for an $\emptyset \neq R \subsetneq V$. 

For the second part, consider some $\emptyset \neq R \subsetneq V$ satisfying $w_x(\delta(R)) \geq \ell$. 
Clearly, there exists a tier $i \in [\ell]$ for which $w_x(\delta(R) \cap E_i) \geq 1$ holds. 
Since $w_x(e) = x_e/(1+q_{\leq i})$ for any tier-$i$ edge $e$, $x(\delta(R) \cap E_i) \geq 1+q_{\leq i}$ holds. 
Now consider an arbitrary vulnerable partition $(J,K)$ of $\delta(R)$. 
By definition, $|J \cap E_{\leq i}| \leq q_{\leq i}$ holds. 
Since residual capacities are nonincreasing over the tiers, we have: 
\[
u(j;R,J) \geq u(i;R,J) \geq (1 + q_{\leq i} - |J \cap E_{\leq i}|)^{-1}.
\]
Therefore, the LHS of \eqref{eq:lpcovering-ext} is at least $x(K \cap E_{\leq i})/(1+q_{\leq i} - |J \cap E_{\leq i}|)$. 
We finish the proof by showing that this ratio is at least $1$:  
\[
x(K \cap E_{\leq i}) = x(\delta(R) \cap E_{\leq i}) - x(J \cap E_{\leq i}) \geq 1+q_{\leq i} - |J \cap E_{\leq i}|,
\]
where we use $x_e \leq 1$ for all $e \in E$. \customqed
\end{proof}

The next result is an extension of \Cref{choiceofJ}. 
For any tuple $\va \in \Rp^\ell$, we use the notation $a_{\leq i} := \sum_{j = 1}^{i} a_j$ with the convention $a_{\leq 0} := 0$. 
The main idea here is that for every fractional solution $x$ and every $\emptyset \neq R \subsetneq V$, checking the validity of the LP constraint~\eqref{eq:lpcovering-ext} can be done by inspecting at most $n^\ell$ vulnerable sets $J \subset \delta(R)$. 
Recall that the instance is assumed to be feasible, so for any $\emptyset \neq R \subsetneq V$, there is a tier $i$ for which $|\delta(R) \cap E_{\leq i}| \geq 1 + q_{\leq i}$ holds. 

Fix some $x \in [0,1]^E$ and some $\emptyset \neq R \subsetneq V$. 
For any tier $i \in [\ell]$, let $L_i$ be an ordered list of edges $e \in \delta(R) \cap E_i$ in nonincreasing order of $x_e$, with ties broken arbitrarily. 
Let $\As \subset \Zp^\ell$ denote the set of size-tuples of all possible choices of vulnerable sets w.r.t. $R$:
\[
\As := \{ \va \in \Zp^{\ell} : \forall \, i \in [\ell], a_i \leq |L_i| \text{ and } a_{\leq i} \leq q_{\leq i}  \}.
\]
Note $|\As| \leq n^{\ell}$. 
For each $\va \in \As$, let $J_{\va}$ denote the (vulnerable) edge set comprising the first $a_i$ edges of $L_i$ for each $i \in [\ell]$. 

\begin{lemma} \label{choiceofJ-ext}
Suppose $x$ satisfies constraint~\eqref{eq:lpcovering-ext} for every vulnerable partition of $\delta(R)$ of the form $J_{\va} \cup K_{\va}$, where $\va \in \As$. 
Then $x$ satisfies the same constraint for every vulnerable $J \subset \delta(R)$. 
Conversely, if constraint~\eqref{eq:lpcovering-ext} is violated for some vulnerable $J \subset \delta(R)$, then the same constraint is violated for $J_{\va}$ satisfying $\va = (|J \cap E_i|)_{i \in [\ell]}$. 
\end{lemma}
\begin{proof}
We prove both parts at once. 
Consider an arbitrary vulnerable $J \subset \delta(R)$ and define $K := \delta(R) \sm J$. 
Consider the $\ell$-tuple $\va = (|J \cap E_1|,\ldots,|J \cap E_\ell|)$. 
Clearly, $\va \in \As$, so $J_{\va}$ is well-defined and vulnerable. 
Let $K_{\va} := \delta(R) \sm J_{\va}$. 
Recall from \eqref{eq:rescapvecfgc} that $u(e;R,J) = u(e^{\prime};R,J_{\va})$ holds for any tier $i \in [\ell]$ and any two tier-$i$ edges $e \in K$ and $e^{\prime} \in K_{\va}$. 
Since $J_{\va}$ comprises the first $a_i$ edges from $L_i$, for each tier $i$, we get the following lower bound on the LHS of \eqref{eq:lpcovering-ext}:
\[
\sum_{e \in K} u(e;R,J) \cdot x_e \geq \sum_{e^{\prime} \in K_{\va}} u(e^{\prime};R,J_{\va}) \cdot x_{e^{\prime}}.
\]
Note that the RHS in the above inequality is precisely the LHS of \eqref{eq:lpcovering-ext} for the vulnerable edge set $J_{\va}$. 
So both conclusions follow. \customqed
\end{proof}

The above two lemmas imply a natural strategy for devising a separation oracle for our LP. 

\begin{lemma} \label{seporacle-ext}
Let $x \in [0,1]^E$. 
In $n^{O(\ell)}$ time, we can determine whether $x$ satisfies constraint~\eqref{eq:lpcovering-ext} for every $\emptyset \neq R \subsetneq V$ and every vulnerable partition $(J,K)$ of $\delta(R)$, or find a violated inequality.
\end{lemma}
\begin{proof}
Let $H_x = (G,w_x)$ denote the capacitated graph derived from $G$ and $x$. 
Using standard tools, we can efficiently compute a minimum cut in $H_x$.  
Let $R^*$ denote one of the shores of this minimum cut, and $\lambda := w_x(\delta(R^*))$. 
If $\lambda < 1$, the first part of \Cref{feasmincut-ext} yields a violated inequality for the trivial vulnerable partition $(\emptyset,\delta(R^*))$. 

Now suppose $\lambda \geq 1$. 
The second part of \Cref{feasmincut-ext} implies that any $\emptyset \neq R \subsetneq V$ with $w_x(\delta(R)) \geq \ell$ cannot produce a violated inequality. 
So it suffices to focus only on those $R$ with $w_x(\delta(R)) < \ell$. 
Define 
\[
\Rs := \{ \emptyset \neq R \subsetneq V : w_x(\delta(R)) < \ell \}.
\]
Note $|\Rs| = O(n^{2\ell})$ (\Cref{nearmincuts}) by our assumption that $\lambda \geq 1$. 
In fact, using \cite{BeidemanCW23}, we can enumerate $\Rs$ in $n^{O(\ell)}$ time. 
Fix some $R \in \Rs$. 
For each tier $i \in [\ell]$, let $L_i$ be an ordered list of edges $e \in \delta(R) \cap E_i$ in nonincreasing order of $x_e$, with ties broken arbitrarily.  
By \Cref{choiceofJ-ext}, if there is a violation in constraint~\eqref{eq:lpcovering-ext} for some vulnerable $J \subset \delta(R)$, then we may take $J = J_{\va}$  with $\va := (|J \cap E_i|)_{i \in [\ell]}$.  
Thus, we either find a violated inequality for this choice of $R$, if one exists, or we have a certificate that there are no violations for any vulnerable $J \subset \delta(R)$.  
The number of choices for $\va$ is at most $\prod_{i \in [\ell]} q_{\leq i} \leq n^{\ell}$, so the overall running time is $n^{O(\ell)}$. \customqed
\end{proof}

In summary, $x$ is feasible for the $\vfgc$ LP if and only if all the following conditions hold: (i)~$x \in [0,1]^E$; (ii)~the minimum cut value in $H_x$ is at least $1$; and (iii)~constraint~\eqref{eq:lpcovering-ext} holds for every $\emptyset \neq R \subsetneq V$ with $w_x(\delta(R)) < \ell$ and every $J_{\va} \subset \delta(R)$ with $\va \in \As$. 
Since the feasible region has bounded facet complexity (and is, in particular, in the box $[0,1]^E$), and a separation oracle running in $n^{O(\ell)}$ time is available, we can use the Ellipsoid method to obtain an optimal solution in time polynomial in $n^{\ell}$ and in the input encoding size; see \cite[Chapter 6]{GLS81}.

\begin{theorem} \label{lpsolver-ext}
Suppose we are given a feasible instance of the $\vfgc$ problem.  
In time polynomial in $n^{\ell}$ and in the input encoding size, we can compute a vector $x \in [0,1]^E$ that is feasible for the linear relaxation of the integer program given by \eqref{eq:lpobj-ext}--\eqref{eq:lpbox-ext} and has $c^\top x := \sum_{e \in E} c_e x_e$ at most \opt.  
\end{theorem}

\subsection{Independent Rounding Algorithm for \texorpdfstring{$\vfgc$}{multi-tier FGC}} \label{indepround-ext}

Our main result in this section is that independent rounding of an LP solution for $\vfgc$ yields an $O(\ell^2 \log n)$-approximation algorithm. 

Let $x \in [0,1]^E$ be a feasible solution to the LP given by \eqref{eq:lpobj-ext}--\eqref{eq:lpbox-ext}. 
Define $y \in [0,1]^E$ as follows: for each edge $e$, $y_e := \min(1,(25 \ell^2 \log n) x_e)$. (Recall that we have $\ell=2$ in $\pqfgc$.)  
Let $F$ be obtained by independently including each edge $e \in E$ in $F$ with probability $y_e$. 
We establish the following theorem using standard concentration inequalities.  

\begin{theorem} \label{rounding-ext} 
With probability at least $1/3$, $F$ is feasible for the given instance of $\vfgc$ and has cost at most $(50 \ell^2 \log n) \cdot c^\top x$. 
\end{theorem}

Applying the above rounding result to an optimal LP solution $x$ obtained from \Cref{lpsolver-ext} yields an $O(\ell^2 \log n)$ approximation algorithm for $\vfgc$ (\Cref{logapx-ext}). 
We can check whether $F \subseteq E$ is a feasible solution for the given $\vfgc$ instance in $n^{O(\ell)}$ time using \Cref{feascheck-ext}. 
Thus, the algorithm can be run repeatedly until a feasible solution with cost $O(\ell^2 \log n) \cdot c^\top x$ is obtained.  
This gives a Las Vegas $O(\ell^2 \log n)$-approximation algorithm with expected running time that is polynomial in $n^{\ell}$ and the input encoding size. 
The proof strategy for \Cref{rounding-ext} is identical to that of \Cref{rounding}, so we skip its proof and instead focus on proving the extensions of \Cref{feaslargevars,badsmallcut,badlargecut}. 

Let $A := \{e \in E : y_e = 1\}$ denote the set of edges (from any tier) that are always included in $F$. 
For any edge $e \notin A$, let $Y_e$ denote the indicator random variable for the event $\{e \in F\}$.
Note $\E[Y_e] = y_e = (25 \ell^2 \log n) x_e < 1$ for any $e \notin A$. 
For any $\emptyset \neq R \subsetneq V$, let $B_R$ denote the indicator random variable for the \emph{bad} event that $\delta_F(R)$ is vulnerable. 
Note that $B_R = 1$ corresponds to a certificate of infeasibility of $F$ (recall \Cref{cutchar-ext}). 
For each tier $i \in [\ell]$, we also define an indicator random variable $B_{R,i}$ for the event $\{ |\delta_F(R) \cap E_{\leq i}| \leq q_{\leq i} \}$ that the random $F$ does not have sufficiently many cut-edges from the $i$ safest tiers. 
The following claim is obvious.

\begin{claim}
We have $B_R = \prod_{i \in [\ell]} B_{R,i}$. 
Therefore, $\Pr[B_R = 1] \leq \min_{i \in [\ell]} \Pr[B_{R,i} = 1]$. 
\end{claim}

The extension of \Cref{feaslargevars} is immediate since $F \supseteq A$ with probability $1$.  

\begin{lemma} \label{feaslargevars-ext}
For any $\emptyset \neq R \subsetneq V$, if $\delta_A(R)$ is not vulnerable, then $\Pr[B_R = 1] = 0$. 
\end{lemma}

Henceforth, we fix an arbitrary $\emptyset \neq R \subsetneq V$ such that $\delta_A(R)$ is vulnerable. 
For a tier $i \in [\ell]$, define 
\begin{equation} \label{eq:rho}
\rho_i := 1+q_{\leq i} - |\delta_A(R) \cap E_{\leq i}|
\end{equation}
to be the minimum number of edges from $(\delta(R) \sm A) \cap E_{\leq i}$ that must be included in $F$ to satisfy the robustness requirement w.r.t. this tier. 
As $\delta_A(R)$ is vulnerable, $\rho_i \geq 1$ for all $i \in [\ell]$. 
Next, we say that a tier $i \in [\ell]$ is \emph{active} if $\rho_i < \rho_j$ for all tiers $j > i$. 
Note that tier $\ell$ is vacuously active. 

\begin{lemma} \label{badsmallcut-ext}
We have $\Pr[B_R = 1] \leq \Pr[B_{R,i} = 1] \leq n^{-5\ell}$ for some active tier $i \in [\ell]$. 
\end{lemma}
\begin{proof}
Since $x$ is a feasible LP solution, constraint~\eqref{eq:lpcovering-ext} is satisfied for all vulnerable partitions $J \cup K$ of $\delta(R)$. 
Like in \Cref{indepround}, we take $J = \delta_A(R)$ and $K = \delta(R) \sm J$. 
We restate \eqref{eq:lpcovering-ext} for convenience using the $\rho$ notation and recalling that the residual capacities (see \eqref{eq:rescapvecfgc}) are uniform within a tier: 
\[
\sum_{i=1}^{\ell} ( \max_{j \geq i} \rho_j^{-1} ) \cdot  x(K \cap E_i) \geq 1.
\]
Since there are $\ell$ tiers, there exist two distinct indices $i,i'$ such that: (a)~$0 \leq i' < i \leq \ell$; (b)~tier $i$ is active but no tier $j \in \{i^{\prime}+1,\ldots,i-1\}$ is active; and 
(c)~we have 
\[
\rho_i^{-1} \cdot x(K \cap (E_{i^{\prime}+1} \cup \ldots \cup E_{i})) \geq 1/\ell.
\] 
As $y_e = (25 \ell^2 \log n) x_e$ for all $e \notin A$, we get: 
\begin{equation} \label{eq:meanforsmallcuts-ext}
y(K \cap (E_{i^{\prime}+1} \cup \ldots \cup E_{i})) \geq (25 \ell \log n) \rho_i =: \mu_i. 
\end{equation} 

Let $Z_i := \sum_{e \in K \cap E_{\leq i}} Y_e$ denote the random number of $K$-edges from the $i$ safest tiers that are included in $F$.   
Note that \eqref{eq:meanforsmallcuts-ext} implies $\E[Z_i] \geq \mu_i$. 
Observe that: 
\begin{align*}
\Pr[B_{R,i} = 1] & = \Pr[|\delta_F(R) \cap E_{\leq i}| \leq q_{\leq i}] \leq \Pr[Z_i < \mu_i / 3] & \tag{recall \eqref{eq:rho} and \eqref{eq:meanforsmallcuts-ext}} \\ 
& \leq \exp(-(5 \ell \log n) \rho_i) \leq n^{-5\ell}.  \tag{\Cref{chernoff}, $\rho_i \geq 1$}
\end{align*}
\noindent To conclude, the bad event associated with $B_R$ occurs with probability at most $n^{-5\ell}$. \customqed
\end{proof}

Our next result is that cuts with large $w_x$ capacity (recall \Cref{eq:edgecap-ext}) have a significantly lower failure probability. 

\begin{lemma} \label{badlargecut-ext}
Suppose that the following capacity lower bound holds for some $\tau \geq 2$: 
\[
w_x(\delta(R)) = \sum_{i \in [\ell]} (1+q_{\leq i})^{-1} \cdot x(\delta(R) \cap E_i) \geq \tau \ell.
\]
Then $\Pr[B_R = 1] \leq n^{-5 \tau \ell}$. 
\end{lemma}
\begin{proof}
By our assumption, there is a tier $i \in [\ell]$ such that $x(\delta(R) \cap E_i) \geq (1+q_{\leq i}) \tau$ holds. 
Since $x \in [0,1]^E$ and $\tau \geq 2$, for the vulnerable $J := \delta_A(R)$ and $K := \delta(R) \sm J$, we have:  
\[
x(K \cap E_i) \geq (1+q_{\leq i}) \tau - x(J \cap E_i) \geq (1+q_{\leq i}) \tau/2.
\]
As $y_e = (25 \ell^2 \log n) x_e$ for all $e \in K$, we get: 
\begin{equation} \label{eq:meanforlargecuts-ext}
y(K \cap E_i) \geq (25 \ell^2 \log n) (1+q_{\leq i}) \tau/2 \geq (25 \ell \log n) (1+q_{\leq i}) \tau =: \mu_i. 
\end{equation} 

Again, let $Z_i := \sum_{e \in K \cap E_{\leq i}} Y_e$. 
Note $\E[Z_i] \geq \mu_i$ by \eqref{eq:meanforlargecuts-ext}. 
Observe that: 
\begin{align*}
\Pr[B_{R,i} = 1] & = \Pr[|\delta_F(R) \cap E_{\leq i}| \leq q_{\leq i}] \leq \Pr[Z_i < \mu_i / 3] & \tag{recall \eqref{eq:meanforlargecuts-ext}} \\ 
& \leq \exp(- (5 \ell \log n)(1+q_{\leq i})\tau) \leq n^{-5\ell\tau}.  \tag{\Cref{chernoff}}
\end{align*}
\noindent To conclude, the bad event associated with $B_R$ occurs with probability at most $n^{-5\ell \tau}$. \customqed
\end{proof}

\section{Conclusions} \label{conclusions}

In this paper, we give a new linear programming relaxation for $\pqfgc$ and obtain a randomized $O(\log n)$-approximation algorithm by independent rounding. 
An immediate open question is whether this algorithm can be derandomized. 
Recall that our algorithm and its analysis follow the template for the randomized $O(\log n)$-approximation algorithm for $\capkecss$, due to Chakrabarty, Chekuri, Khanna, and Korula \cite{CCKK15}. 
Only recently, Bansal, Cheriyan, Khanna, and Simmons \cite{BCKS25} were able to obtain a different deterministic $O(\log k)$-approximation algorithm for $\capkecss$ by using an LP-based deterministic constant-factor approximation algorithm for the CoverSmallCuts problem; however, since the edge capacities arising in our LP depend on the cut, we are currently unable to adapt their ideas to our setting. 

A central question is whether $\pqfgc$ can be approximated within factors independent of the size of the graph (i.e., $n$). 
Another question is whether the integrality gap of our LP relaxation can be bounded by a function that depends only on $p$ and $q$, or even by a universal constant? 
The integrality gap of our LP is lower bounded by the integrality gap of the covering LP arising in the augmentation version of the problem. 
Here, the augmentation version refers to a special case of $\pqfgc$ where the set $F_0$ comprising the zero-cost edges already forms a feasible $(p,q-1)$-$\fgc$ solution. 
The ``vulnerable'' cuts $R$ that need to be strengthened satisfy $|\delta_{F_0}(R) \cap \safe| < p$ and $|\delta_{F_0}(R)| = p+q-1$. 
Note that the knapsack cover inequalities for the augmentation LP are standard covering constraints, since there is no qualitative distinction between safe and unsafe edges in the augmentation problem.  

In the context of network design, a standard and powerful technique is iterative rounding, first introduced by Jain for survivable network design \cite{Jain01}. 
There are two main obstacles to using this approach in our linear program. 
First, the demand function does not satisfy skew submodularity or other properties that enable uncrossing. 
Although the LP can be solved efficiently, it requires checking all near-minimum cuts $R$ and $O(m)$ edge sets $J$ for each $R$. 
Second, iterative rounding is typically applicable to covering and packing LPs with uniform coefficients for each variable; in our constraints, the coefficients of the variables depend on the choice of $J$.

\paragraph{Acknowledgement}
The authors thank Joseph Cheriyan for discussions on the topic that were the main inspiration for this work.

\printbibliography

\end{document}